# Predicting orientation-dependent plastic susceptibility from static structure in amorphous solids via deep learning

Zhao Fan* and Evan Ma

*Department of Materials Science and Engineering, Johns Hopkins University, Baltimore, MD 21218, United States*

* zfan7@jhu.edu

## Abstract

It has been a long-standing materials science challenge to establish structure-property relations in amorphous solids. Here we introduce a rotation-variant local structure representation that enables different predictions for different loading orientations, which is found essential for high-fidelity prediction of the propensity for stress-driven shear transformations. This novel structure representation, when combined with convolutional neural network (CNN), a powerful deep learning algorithm, leads to unprecedented accuracy for identifying atoms with high propensity for shear transformations (i.e., plastic susceptibility), solely from the static structure – the spatial atomic positions – in both two- and three-dimensional model glasses. The data-driven models trained on samples at one composition and a given processing history are found transferrable to glass samples with different processing histories or at different compositions in the same alloy system. Our analysis of the new structure representation also provides valuable insight into key atomic packing features that influence the local mechanical response and its anisotropy in glasses.

# Introduction

The mechanical response of a crystalline metal to applied stresses can be quantitatively explained by monitoring the evolution of dislocations[1, 2] in the lattice. In stark contrast, amorphous metals do not have such well-defined structural defects, owing to the absence of long-range atomic packing order. As a result, in the static structure of a glass, it remains a grand challenge to identify *a priori* local "defects" that are vulnerable to rearrangements, even when the positions of all atoms are fully known[3-7]. Over the years, a number of physical parameters have been used to serve as indicators[8-27] to forecast local regions as fertile sites for shear transformations. Data-driven models[28-31] have been put forward as well for the same goal. However, all these attempts have only achieved moderate success: the correlations with the plastic susceptibility of particles/atoms, i.e. the propensity of each particle to experience plastic rearrangement, have not been sufficiently strong to predict structure-property relations in glassy solids.

A critical reason for this status quo is that the scalar or rotation-invariant quantities invoked so far are inherently inadequate in capturing the anisotropic response of a local configuration. A given local environment of an atom (particle) can respond quite differently when the externally applied global force is imposed along different orientations[32-35]. Recently, additional indicators have been invoked to take into account anisotropy in predicting the propensity for plastic activity in two-dimensional (2D) model glasses[33-35]. But it remains unclear how well these new indicators fare in three-dimensional (3D) amorphous solids. Furthermore, these indicators are physical quantities that need to be evaluated with known interparticle interactions, every time the structure changes. In other words, one cannot simply monitor the atomic positions alone, to explain property changes. In terms of physical insight, it is difficult, if not impossible, to use these methods to directly identify the critical structure features (i.e., the atomic environment *per se*) responsible for the anisotropic local mechanical response. This has been a puzzle existing for at least more than two decades since Falk and Langer pointed it out in 1998[32].

Recently, deep convolutional neural networks (CNN) have enabled revolutionary breakthroughs in the field of computer vision and pattern recognition[36-39]. Inspired by such successes, here we explore the potential of this deep learning approach via converting local configurations into images and constructing large-scale training datasets (up to ~20 million images) from extensive atomic simulations. More importantly, we also devise a rotation-variant structure representation, to capture the anisotropic responses of a local configuration and the intrinsic sensitivity to the loading orientation. We will demonstrate that this combination achieves unprecedented accuracy when it is used to separate out particles with high plastic susceptibility from the rest in the sample, unveiling the power of local static structure for predicting orientation-dependent plastic events in both 2D and 3D glasses. In terms of new insight, by identifying a clear difference in the new atomic structure representation for contrasting particles, i.e., those with high versus low plastic susceptibility, we have answered, from the structural perspective, the puzzle as to why the same local configuration could respond very differently to external loading applied in different orientations.

## Atomic structure representation and model architecture

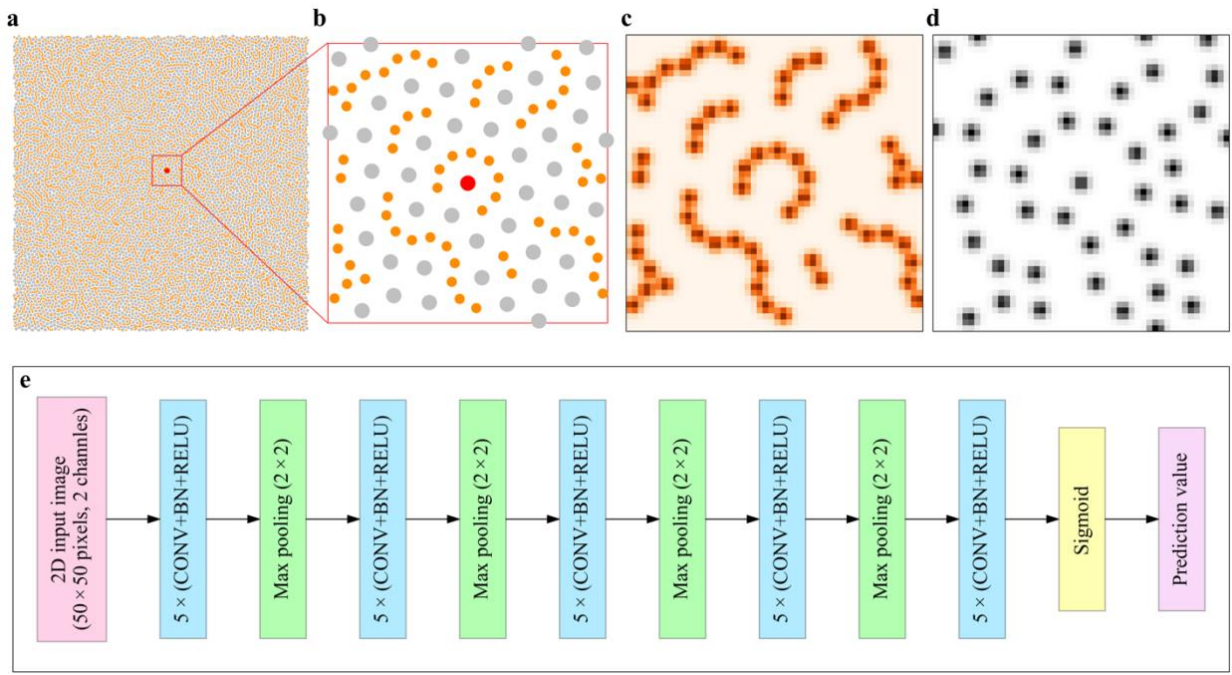

**Fig. 1. Atomic structure representation and model architecture for 2D glasses**. (**a**) Snapshot of a 2D model glass. Orange and silver circles represent small (*S*) and large (*L*) particles, respectively. (**b**) A close-up showing a local configuration around the red particle in (**a**). (**c**) and (**d**) Corresponding spatial density map (SDM) of the *S* and *L* particles, respectively. (**e**) Architecture of the convolutional neural network (CNN) model used for 2D glasses, which contains 25 (5 in each blue box) 2D convolutional (conv.) layers. A 2D max-pooling layer is periodically inserted in-between the successive conv. layers. The last conv. layer is followed by the output layer which is a sigmoid neuron. The corresponding architecture of the CNN model for our 3D model glasses is shown in Supplementary Fig. 3.

We have developed CNN models in both a 2D binary Lennard-Jones (L-J) glass system[40] and a 3D binary embedded-atom-method (EAM)-based Cu-Zr metallic glass system[41]. To represent the local packing environment of each particle $i$, we use a new multi-dimensional structural function, termed a Gaussian-weighted spatial density map (SDM). For the 2D model, the SDM of each particle $i$ is defined as

$$\Xi_i(y, x, \beta) = \sum \exp\left(-\frac{(r_{ij,x}-x)^2+(r_{ij,y}-y)^2}{2\Delta^2}\right), \qquad [1]$$

where the summation is performed over all particles satisfying these conditions: species$(j) \in \beta$; $|r_{ij,x}| < r_c$ and $|r_{ij,y}| < r_c$, $r_{ij,x}$ and $r_{ij,y}$ are the components along $x$ and $y$ dimensions, respectively, of $\boldsymbol{r}_{i,j}$, which is the vector connecting the central particle $i$ with particle $j$ of species $\beta$ within a cutoff $r_c$ (particle $j$ may be the central particle when the species of the central particle is that under consideration in the channel). Here $x, y \in [-r_c + 0.5\Delta, r_c - 0.5\Delta]$ with a constant increment of $\Delta$, and $\beta \in \{0, 1\}$, where 0 represents small ($S$) particles, and 1 large ($L$) ones. We set $r_c = 5\,\sigma$ and $\Delta = 0.2\,\sigma$ after trials using many different values (see Supplementary Fig. 1a, Supplementary Note 1 and Supplementary Note 2, and $\sigma$ is the characteristic length scale parameter in the potential, see details in *Methods*).

Similarly, for 3D glass systems, the SDM of each atom $i$ is defined as

$$\Xi_i(z, y, x, \beta) = \sum \exp\left(-\frac{(r_{ij,x}-x)^2+(r_{ij,y}-y)^2+(r_{ij,z}-z)^2}{2\Delta^2}\right), \qquad [2]$$

here the summation is performed over all particles satisfying these conditions: species$(j) \in \beta$; $|r_{ij,x}| < r_c$; $|r_{ij,y}| < r_c$ and $|r_{ij,z}| < r_c$, and $\beta = 0$ or 1, to represent Zr and Cu atoms, respectively. Using SDM to represent local configurations can be viewed as projecting the local configurations to 2D (or 3D) grids for different species, which results in a multi-dimensional numerical array, equivalent to a 2D (or 3D) image containing $(2\,r_c/\Delta)^2$ (or $(2\,r_c/\Delta)^3$) pixels. And each pixel has channels equal to the number of components in the sample. These images can be used directly as input into the CNN model. For an adequately large $r_c$, a sufficient number of surrounding particles would be included. And, when $\Delta$ is small enough, any tiny variation of

particle positions in the local configuration would lead to a corresponding variation in SDM. Therefore, the SDM can faithfully and accurately represent the topological structure feature of a local configuration. Obviously, when a larger $r_c$ is used, an image would contain structural information across a larger range; and a smaller Δ means a higher resolution of resultant images; the difference between two very similar local configurations would be more pronounced in the images. Although a larger $r_c$ and a smaller Δ are good for more faithful and accurate representation of the topological structure feature and elevated predictive performance, these also mean larger size of each image, requiring more computational resource and memory to generate, store and use these images. In this work, we will choose optimized values of both $r_c$ and Δ for different glass systems after trying many different values. See more discussion about the optimization of both $r_c$ and Δ in Supplementary Note 1 and Supplementary Note 2. In addition, the contribution to the SDM from different species are separated into different channels, taking into account the influence arising from chemical affinity. Fig. 1a shows a typical 2D model glass, with a close-up view (a local configuration) in Fig. 1b. The SDM of *S* and *L* particles corresponding to Fig. 1b are shown in Fig. 1c and 1d, respectively, where each image contains 50 × 50 pixels. The pixels with high intensity (darker color) correspond to the locations of particles. Obviously, the SDM will be different when the local configuration is rotated. Before computing SDMs, we will use different coordinate systems or rotate configurations for different loading orientations such that the shear direction and the normal direction of the shear plane are parallel to X and Y axes, respectively, of the chosen coordinate system (several specific examples are given in Supplementary Fig. 2 for choosing appropriate coordinate systems for different loading protocols). Therefore, the SDM is rotation-variant, which is the key feature we need for predicting orientation-dependent plastic events in amorphous solids. As will be shown later in Fig. 4, the rotation-dependent SDM is very sensitive to small changes in sample-loading alignment. The SDM can also be used to predict rotation-invariant properties of crystalline materials via data augmentation[36], i.e. arbitrarily rotating images each time they are fed into the training.

We then applied the CNN method[39] to each input image to predict the propensity of each particle to experience plastic rearrangement, at different global shear strains when the sample is loaded along a specific orientation. Fig. 1e illustrates the architecture of the CNN model for the 2D glasses. Specifically, each CNN model contained 25 convolutional (conv.) layers, where filters with a small receptive field of 3 × 3 were used. Zero-padding was involved in all but the fifth and last conv. layers. Batch normalization (BN)[42] was adopted right after each convolution and before activation with the rectification (ReLU) non-linearity[36]. A max-pooling layer is periodically inserted in-between the 25 successive conv. layers. Max-pooling is performed over a 2 × 2 pixel window, with a stride of 2. The number of filters is 20 in the first five conv. layers and then doubled after each max-pooling layer. And the last conv. layer is directly followed by the output layer, which is a sigmoid neuron as we are performing binary classification tasks.

For 3D glasses, $r_c = 9.6$ Å and $\Delta = 0.6$ Å were found to produce optimized results (see Supplementary Fig. 1**b**, Supplementary Note 1 and Supplementary Note 2), and the resultant input images contain 32 × 32 × 32 pixels. The corresponding CNN model contains 20 conv. layers and 3 max-pooling layers, and both convolution and max-pooling were performed in 3D space (the architecture of the CNN for our 3D glasses is presented in Supplementary Fig. 3). The number of filters in each 3D conv. layer was 60. To accelerate training and avoid shortage of memory, we did not double the filter number after max-pooling. This is not expected to influence significantly the training results, based on our experience with the 2D glasses for which doubling the filter number increased the validation accuracy by less than 1%. For tutorials about deep learning, the readers are referred to Refs[36-39].

## Structural differences responsible for the anisotropy of local mechanical response

In the literature, most authors quote "free volume" as the structural feature controlling the (local) susceptibility to rearrangement in amorphous solids[8, 43]. However, as will be shown later in this work, no obvious correlation is found between the plastic susceptibility and local excess volume. There have also been recent observations that atoms with fewer neighboring particles in the nearest neighbor shell, but more in the troughs in between the shells, would be more flexible[28, 30]. However, this does not explain why the same local configuration responds differently to applied global force along different directions. Now that we have converted local configurations into images, let us first examine, from this perspective, if we could identify a critical structure feature responsible for the anisotropic mechanical response. To this end, we seek valuable insight by contrasting the SDMs of particles with extremely high plastic propensity (i.e., the tail end of the high-susceptibility side) versus those on the other end.

We first sheared 5,000 2D glass samples to a shear strain of 3.0% using the athermal quasi static (AQS) method[44, 45], such that we can identify all of the particles with extreme (the highest or the lowest 0.5%) non-affine squared displacements ($D^2_{min}$)[32], for each of the four different loading protocols (positive and negative simple shear, and positive and negative pure shear, see Supplementary Fig. 2 and *Methods*). In this work, following the pioneering work by Falk and Langer[32] as well as recent machine learning attempts[28, 29], we assume that an atom has a high plastic susceptibility if it shows a large $D^2_{min}$ (the position of the atom is then a fertile site for shear transformations). The plastic susceptibility is a scalar quantity, but exhibits a different value each time the sample is globally loaded along a different orientation. We then went back to the original configurations before deformation, and calculated the SDMs for these two groups of extreme particles that we want to compare. For each group an average SDM was obtained, representing the averaged environment of a representative particle, as shown in the Supplementary Fig. 4 and Supplementary Fig. 5. It is interesting to observe that for the highest 0.5% $D^2_{min}$ (see the 1st and 3rd rows) particle, hereafter referred to as "fertile site", its neighboring shells are more diffuse (wider), and the intensity in the same shell is non-uniform. For the other extreme, the lowest $D^2_{min}$ particle (shown in the 2nd and 4th rows), each shell is sharper with more uniform intensity. In other words, the images do reveal structural differences between particles with high versus low plastic susceptibility.

This difference in local environment can be appreciated by presenting the SDM in another way, as presented in Fig. 2a-h from the standpoint of each *S* particle (at center). Fig. 2a-d are for its *S* neighbors in the surrounding. In each panel, we subtract the SDM for the center *S* particles with the lowest 0.5% $D^2_{min}$ (the 2nd row in Supplementary Fig. 4) from those with the top 0.5% $D^2_{min}$ (the 1st row in Supplementary Fig. 4). Similarly, Fig. 2e-h are for the *L* neighbors surrounding an *S* particle, each panel resulting from the 3rd row in Supplementary Fig. 4 *minus* the 4th row. The corresponding SDM difference in the *L*-centric view are provided in Supplementary Fig. 6a-h. As can be seen from these maps, for "fertile sites" the surrounding shells are ellipses. The long axis is parallel to the macroscopic elongation direction **ζ** - the white dashed line in the map, which is the sum of the unit vector $\hat{\iota}$ (effective shear direction) and $\hat{\eta}$ (normal to the effective shear plane), and the short axis is perpendicular to $\zeta$. In contrast, for particles on the other end of the plastic susceptibility spectrum, their neighboring shells are circles.

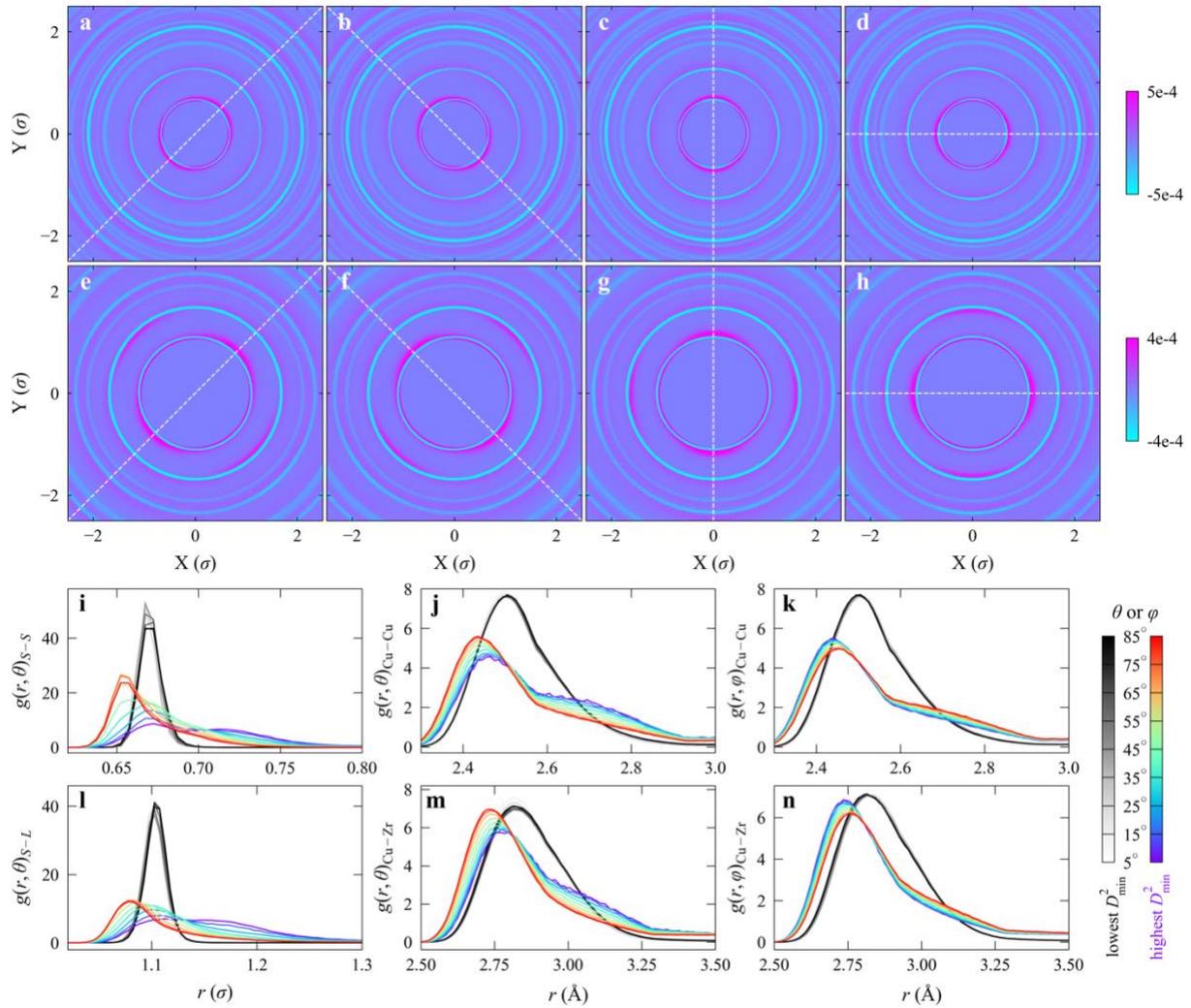

**Fig. 2. Local structural difference between particles with the highest and those with the lowest plastic susceptibility**. (**a-d**) Difference in SDMs for *S* particles with extremely high verses low plastic susceptibility in the 2D glass model. See colored scale bar on the right, for the magnitude of the difference, which is calculated using the first row in Supplementary Fig. 4 *minus* the second row. Each panel maps out the spatial density of surrounding *S* particles, for one of the four different loading protocols (from left to right: positive simple shear, negative simple shear, positive pure shear and negative pure shear). The white dashed line on each map represents the macroscopic elongation direction, $\zeta$. The corresponding maps for the *L* particles surrounding the *S* are shown in (**e-h**). (**i**) and (**l**) show orientational partial (*S*-centered) pair correlation function $g_\zeta(r,\theta)$, for the *S*-*S* and *S*-*L* correlation, respectively. Each $g_\zeta(r,\theta)$ curve is an average, over all samples and all loading orientations. For the 3D $Cu_{50}Zr_{50}$ glass, we only show the

orientational pair correlation functions for atoms that showed extreme $D^2_{min}$ upon straining to 5%. (**j**) and (**m**) show Cu-centric $g_\zeta(r,\theta)$ curves, for Cu-Cu and Cu-Zr, respectively. And (**k**) and (**n**) show Cu-centric $g_\zeta(r,\varphi)$ curves, for Cu-Cu and Cu-Zr, respectively. The corresponding figures for *L* particle as center (2D glass) and Zr atom as center (3D glass) are shown in Supplementary Fig. 6.

To further understand the structure difference, we also used an orientational pair correlation function, $g_\zeta(r,\theta)$, to characterize the local atomic packing environment. The $g_\zeta(r,\theta)$ for a single particle *i* is defined as:

$$g_{\zeta,i}(r,\theta)_\beta = \frac{1}{\kappa\rho f_\beta} \sum_{j=1}^{N_\beta}\left[\delta(r-r_{ij})\delta\left(\theta-\theta_{\zeta,r_{ij}}\right)\right] \quad [3]$$

where $\theta_{\zeta,r_{ij}} = \arccos(|\hat{r}_{ij}\cdot\hat{\zeta}|)$ is the angle between the vectors $\boldsymbol{\zeta}$ and $\boldsymbol{r}_{ij}$ and has its values in the range of $[0, \pi/2]$, $r_{ij} = |r_{ij}|$, $\rho$ is the number density of atoms, and $f_\beta$ is the composition fraction of species $\beta$ in a sample. For 2D samples, $\kappa = 4r\Delta r\Delta\theta$, and for 3D samples, $\kappa = 4\pi r^2\Delta r[\cos(\theta - 0.5\Delta\theta) - \cos(\theta + 0.5\Delta\theta)]$. The $g_\zeta(r,\theta)$ measures the line density of particles/atoms in the region with radial distance of *r* from the central atom *i*, oriented relative to elongation direction $\boldsymbol{\zeta}$ at angle $\theta$ (this can be compared with the usual pair correlation function, which can be regarded as the average of orientational correlations over all $\theta$). Again, our goal is to probe the local environment of *S* (or *L*) particles identified to exhibit extreme $D^2_{min}$ in 3.0%-strained 2D samples. In order to accomplish this we went back to the original unsheared sample to calculate the $g_{\zeta,i}(r,\theta)$ for these particles of interest. This curve is then averaged over all such *S* (or *L*) particles from 5,000 different samples for a given protocol (each sample was sheared 4 times, with the different protocols). The $g_\zeta(r,\theta)$ plots are shown in Supplementary Fig. 7 for *S* particles and Supplementary Fig. 8 for *L* particles. The $g_\zeta(r,\theta)$ curves obtained for particles identified in the various loading protocols show little difference; e.g., (**a-d**) in the Supplementary Fig. 7 are very similar to one another; they were hence averaged to the one shown in Fig. 2i.

As seen from these plots, for particles with low plastic susceptibility, the grey curves of $g_\zeta(r,\theta)$ for different $\theta$ almost overlap on top of each other, which is consistent with the observation that around each of such particles the neighbors tend to form circularly symmetric shells. In other words, a particle with low plastic susceptibility tends to have almost the same bond length with its $n^{th}$ neighbors. In stark contrast, for fertile-site particles, the curves of $g_\zeta(r,\theta)$ at different $\theta$ are quite different. Specifically, their neighboring particles with $\theta$ close to 0 have longer bond lengths, whereas those with $\theta$ closer to $\pi/2$ have shorter bond lengths, see **Fig. 2i** and **l**. Around $L$ particles with high plastic susceptibility, the distribution of $S$ neighbors (see Supplementary Fig. 6i) is similar to that observed in **Fig. 2i** and **l**, while the distribution of $L$ neighbors changes with $\theta$ mainly in the peak intensity (see Supplementary Fig. 6l), also indicating the asymmetric packing in the surrounding. Note that other peaks at larger radial distance behave similarly, so only the first peaks are compared in the figures. We carried out similar $g_\zeta(r,\theta)$ analysis for 3D Cu$_{50}$Zr$_{50}$ model glasses. As shown in Fig. 2 and Supplementary Fig. 6 **j** and **m**, the structural features are similar to those in 2D glasses.

The analysis based on $g_\zeta(r,\theta)$ unveils that a local configuration vulnerable to plastic rearrangement has its longest axis of the elliptical neighboring shells parallel to the elongation direction of global loading. This picks out the softest direction of a 2D local configuration. But for a 3D local configuration under this same condition, there is still one more degree of freedom: the local configuration can rotate arbitrarily around the elongation direction. Additional conditions are therefore needed to identify the softest direction of 3D local configurations, to take into account this angle of rotation. To this end, we employ another orientational pair correlation function, $g_{\zeta,i}(r,\varphi)$,

$$g_{\zeta,i}(r,\varphi)_\beta = \frac{1}{\kappa \rho f_\beta} \sum_{j=1}^{N_\beta} [\delta(r-r_{ij})\delta(\varphi-\varphi_j)].  \qquad [4]$$

Here, we shift the elongation vector $\boldsymbol{\zeta}$ such that it goes through the central atom $i$ of a local configuration and then define a new vector $\boldsymbol{\chi}_j$ for each surrounding atom $j$ of the local configuration. The $\boldsymbol{\chi}_j$ is perpendicular to $\boldsymbol{\zeta}$ and passes through the surrounding atom $j$ of interest. Then $\varphi_j$ is the angle between $\boldsymbol{\chi}_j$ and the plane ($\boldsymbol{\iota}\times\boldsymbol{\eta}$) defined by the shear direction vector ($\boldsymbol{\iota}$) and the normal vector of shear plane ($\boldsymbol{\eta}$), for each surrounding atom $j$. Thus, the $g_{\zeta,i}(r,\varphi)_\beta$ depicts neighboring shells which pass through the vector $\boldsymbol{\zeta}$ but with different $\varphi$ relative to the plane of $\boldsymbol{\iota}\times\boldsymbol{\eta}$. When $\varphi=0$, the shells are in the plane of $\boldsymbol{\iota}\times\boldsymbol{\eta}$, and when $\varphi=\pi/2$, they are perpendicular to the plane of $\boldsymbol{\iota}\times\boldsymbol{\eta}$. Here $\kappa=8r^2\Delta r\Delta\varphi$, different from that for $g_\zeta(r,\theta)$ in equation [3].

Figs. 2**k** and **n** contrast the partial $g_{\zeta,i}(r,\varphi)$ of the Cu atoms having the lowest 0.5% $D^2_{min}$ with those having the 0.5% highest for Cu-Cu and Cu-Zr. The corresponding Zr-centric $g_{\zeta,i}(r,\varphi)$ for Zr-Cu and Zr-Zr are shown in Supplementary Figs. 6**k** and **n**, respectively. These extreme particles were identified in the 1,500 $Cu_{50}Zr_{50}$ glasses deformed to 5% strain. For the lowest 0.5% $D^2_{min}$, all partial $g_{\zeta,i}(r,\varphi)$ at different $\varphi$ also overlap with one another, in conjunction with the overlapping partial $g_{\zeta,i}(r,\theta)$ at different $\theta$ shown in Figs. 2**j** and **m** and Supplementary Figs. 6**j** and **m**. These unequivocally confirm that the neighboring shells of particles with lowest plastic susceptibility exhibit spherical symmetry, for the 3D glasses. In other words, the bond lengths in a given shell are very close and the shell is uniformly populated. In contrast, for the fertile sites with the highest 0.5% $D^2_{min}$, the surrounding particles inside the plane of $\iota \times \eta$ tend to have shorter bond lengths, as seen in all the $g_{\zeta,i}(r,\varphi)$ at small $\varphi$. Combining this with the insight from Figs. 2**j** and **m** and Supplementary Figs. 6**j** and **m**, a 3D local configuration is expected to emerge as a fertile site when i) the degree of asymmetry of its atomic packing is sufficiently large; ii) its longest axis is parallel to the elongation direction of the global loading; iii) its orientation around the elongation direction axis is in such a way that the neighboring shell inside the plane of $\iota \times \eta$ coincides with the ellipse having the largest aspect ratio. Putting it another way, in the plane of $\iota \times \eta$, the larger the ratio of the bond length parallel to the elongation direction to that parallel to the contraction direction (the elongation and contraction directions are perpendicular to each other), the higher the susceptibility to plastic rearrangement.

We emphasize here that these features above provide new insight into the difference in local structure, between particles with high versus low plastic susceptibility. To recapitulate, in both the 2D and 3D glasses, for particles with low plastic susceptibility, their neighbors tend to form circular shells with almost uniform intensity (peak height) at different $\theta$ and $\varphi$. In other words, the bond lengths are very similar for each shell, in which the neighbors uniformly distribute. In contrast, packing tends to be more asymmetric surrounding fertile sites, the neighboring shells are ellipses, each with non-uniform particle distribution. And for a fertile site to be activated upon a particular mechanical loading, the longest and shortest axis of neighboring shells should be parallel to the elongation and contraction directions, respectively. In other words, a fertile site is the most prone to be activated when its surrounding packing along the elongation direction is the loosest and packing along the contraction direction is the densest. These structural differences are responsible for the anisotropy of local mechanical response in amorphous solids.

With these insights in mind, we expect that the mechanical response of a local configuration will not show much difference when one mirrors the local configuration around the plane perpendicular to the plane of $\iota \times \eta$ and containing the vector $\zeta$ or the vector perpendicular to $\zeta$, or the plane of $\iota \times \eta$ (see details on these mirror operations in Supplementary Fig. 9), as these mirrors will not change the packing density of local configurations along the elongation or contraction direction. To verify this, we loaded 3 such mirrored configurations of a 2D L-J sample (in both simple and pure shear along a constant loading orientation), and 7 mirrored $Cu_{50}Zr_{50}$ configurations (in simple shear along a constant loading orientation). As expected, $D^2_{min}$ of all particles in those mirrored samples are indeed almost identical to those in the corresponding original samples, when the strain is well below global yielding, as shown in Supplementary Fig. 10. In training our CNN model (next section), we will augment our training data with these mirror-symmetries. This was found to improve accuracy by ~2%, lending further support to our insight above.

## Training procedure and results

The insight in Fig. 2 is useful for our understanding, as it provides a general picture of the main structural features/difference. But there are also other subtle but nontrivial features to be taken into account, when a concrete decision is to be made regarding whether a particular particle has high (or low) susceptibility. As will be demonstrated in the work that follows, deep learning is able to include additional and more subtle information so as to provide such predictive power. And the consistency between the CNN prediction with our expectation based on the insight in Fig. 2 will also be confirmed at the end of this sub-section. To construct big datasets for the CNN models, we prepared 2D L-J (or 3D $Cu_{50}Zr_{50}$) glasses with constant a cooling time of $10^6$ $t_0$ (or effective quench rate of $10^{10}$ K/s) from the corresponding well-equilibrium liquid. 5,000 (1,500) samples were made, each containing 10,000 (32,000) particles. Each sample was then deformed under 4 (24) different loading protocols (see more details about sample preparation and deformation in *Methods* section). These 2D (3D) glass models were divided into training, validation and test datasets, each containing 4,900 (1,480), 50 (10) and 50 (10) samples, respectively. Since we are interested in identifying particles which will experience extreme (large) plastic rearrangement upon loading, the first step is to do a binary classification task. We will explore the relations between initial static structure and plastic responses at different degrees of strain before the yielding point. Thus, we constructed different datasets for the plastic responses at different degrees of strain. For a specific strain, a particle $i$ is labeled $y_i = 1$ if its $D^2_{min}$ is higher than a specific threshold and $y_i = 0$ otherwise. Here we use $f_{thres}$ to set the $D^2_{min}$ threshold at a given shear strain. For example, $f_{thres} = 0.5\%$ means that we label particles in the top 0.5% $D^2_{min}$ group as $y_i=1$ and the remainder $y_i=0$. The shear stress-strain curves and cumulative distribution function (CDF) of $D^2_{min}$ are shown in Supplementary Fig. 11 and Supplementary Fig. 12, respectively, for the 2D and 3D glasses. We will mainly show the predictive performance obtained with $f_{thres} =0.5\%$, after trying many different values of $f_{thres}$ (see more discussions and results in Supplementary Note 3 and Supplementary Fig. 13). Our training, validation and test datasets were balanced, as in each dataset we included all the $y_i = 1$ particles in each sample deformed under each specific deformation protocol, with the same number of $y_i = 0$ particles selected at random from the rest. In contrast to previous works[28, 31], we will train a single model for all species in a glass system to improve the user-friendliness of machine learning models; this does not sacrifice much predictive power, as verified in Supplementary Fig. 14 and related discussion in Supplementary Note 4. Note again that the SDM were constructed for particles in the initial undeformed configurations; it is just that these particles were selected based on their $D^2_{min}$ response upon straining.

In performing binary classification task, the most commonly used metric to measure the predictive performance is "accuracy", defined as the fraction of the number of images (local configurations) predicted correctly divided by the total number of images in a dataset. The image is predicted correctly, when the actual label $y_i$ (1 or 0, determined by the value of $D^2_{min}$ from simulations) of an image is identical to the label predicted by the machine learning model. Since each of our datasets is balanced, i.e., the number of images with actual $y_i = 0$ is equal to that of images with actual $y_i = 1$, the lowest value of accuracy would therefore be 50%. We confirmed that, for each trained model, the validation accuracy and test accuracy are practically the same, suggesting that the trained models did not bias our validation datasets despite of the hyperparameters involved, such as CNN architecture, $r_c$ and $\Delta$. We therefore only report the validation accuracy (or simply "accuracy") in the following.

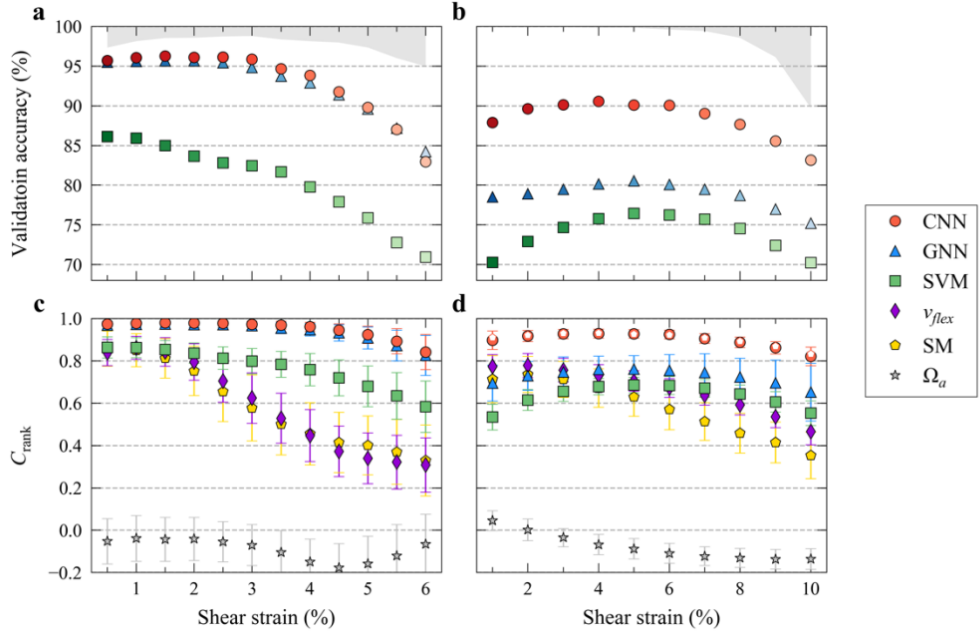

**Fig. 3. Predictive power enabled through deep learning**. (**a**) Accuracy achieved using three different machine learning methods (CNN, GNN and SVM) at different shear strains using a constant $f_{thres}$ of 0.5% for 2D L-J glasses. (**b**) shows the corresponding results for 3D $Cu_{50}Zr_{50}$ glasses. The lower edge of the region shaded grey is the upper bound, i.e., the ceiling predicted using labels based on new $D^2_{min}$ after conducting another deformation simulation on samples belonging to the test dataset (see more details in *Methods* section). (**c**) The cumulative rank correlation ($C_{rank}$, see text for definitions) between real local plastic response in a sample along a give loading orientation and the plastic susceptibility predicted via each of the six routes, including the three machine learning models and the other three based on physical parameters, i.e., flexibility volume ($v_{flex}$), soft modes (SM), atomic volume ($\Omega_a$). The corresponding results for the 3D $Cu_{50}Z_{50}$ glasses are shown in (**d**). The $C_{rank}$ value are the average for 50 (10) different samples loaded in 4 (24) different loading orientations for 2D (3D) glasses, with the error bar marking the standard deviation. For the 3D glasses, each sample contained 32,000 atoms, although the soft mode analysis was conducted using smaller samples each containing 10,000 atoms. For a fair comparison, CNN was also applied on these smaller samples, as shown with open pentagons (i.e., white pentagons overlapping with red full circles) in (**d**).

Fig. 3**a** and **b**, for 2D and 3D glasses, respectively, show the predictive accuracy achieved on validation datasets with our CNN method, the graph neural network (GNN)[31] method and linear support vector machine (SVM)[28] method, when differentiating particles/atoms with $y_i = 1$ from those with $y_i = 0$ at different strains with a constant $f_{thres} = 0.5\%$. As seen from Fig. 3a for the 2D case, CNN exceeds SVM by far in performance, over the entire strain range we considered; the highest accuracy achieved was as high as 96.27%. This accuracy is very close to the ceiling (see *Methods* section for details about the upper bound). Several advantages of CNN make it obviously

superior to SVM: i) CNN can capture more complicated relations between static structure and plastic activity compared to linear SVM, as the latter is merely a linear machine learning algorithm; ii) SVM loses some subtle but critical structural information during the conversion from atomic positions to the input features; iii) most importantly, for a given local configuration in different loading directions, we constructed images with different coordinate systems (in which the X axis is parallel to the shear direction ($\iota$) and the Y axis is the normal direction of the shear plane ($\eta$). This enabled orientation-dependent predictions, whereas the input features to SVM are rotation-invariant, leading always to the same prediction. We observe that GNN produces slightly lower accuracy compared to the accuracy of our CNN, as shown in Fig. 3a. This is not surprising, as GNN used atomic positions directly to construct graphs as input, and in training GNN models we rotated edge vectors for different loading directions, in contrast to the procedure from Ref.[31] (see details in *Methods*). However, the advantage of CNN over GNN is more obvious for larger $f_{thres}$, see Supplementary Fig. 13, and the graph input to GNN cannot be used to provide the insights gained from Fig. 2 based on images, the input to CNN.

When dealing with 3D $Cu_{50}Zr_{50}$ glasses, the GNN method is also superior to SVM, consistent with the observations in Ref.[31]. However, we found that our CNN, as an algorithm with state-of-the-art learning capability, is far more powerful than GNN. This is shown in Fig. 3b, over the entire strain range. The highest accuracy achieved with CNN is above 90%, in the strain range from 3 to 6%. Even at a strain of 10% where global yielding occurs (see Supplementary Fig. 11b), the accuracy is still > 83%. This suggests our CNN model may be useful in predicting the location where shear band initiates. This point is beyond the scope of the current work and will be discussed elsewhere. We believe that the CNN method can achieve higher accuracy, specially for 3D glasses, if we use deeper networks. But it will require more computational resources, see more discussion in Supplementary Note 1. Our main purpose here is to demonstrate that the CNN is indeed a powerful method to meet the challenge we are dealing with.

One would be naturally curious as to how well our CNN models fare, when compared with the prediction based on a correlation between the propensity for plastic rearrangements of each particle and some previously used physical indicators, such as the participation ratio in soft modes (SM)[17,18], flexibility volume ($v_{flex}$)[19], and atomic volume ($\Omega_a$). To allow this comparison, here we evaluate the class probability from CNN and GNN, and the distance to the separation boundary predicted by SVM, to denote the predicted plastic susceptibility of each particle in the samples constructed for the test dataset but never involved in the training dataset. Take CNN as an example, the class probability for each input image is a numerical value in the range of (0, 1) provided by the learning algorithm, to represent the probability of having a label of $y_i=1$ for the input object. For binary classifications, class probability greater than 0.5 predicts $y_i=1$ and $y_i=0$ otherwise. Then the plastic susceptibility of each particle can be predicted via one of the six approaches (the three machine learning methods and three physical parameters). Similar as in Ref.[22], we define a cumulative rank correlation ($C_{rank}$) to compare the power of these six routes in predicting the particles which will have top 0.5% $D^2_{min}$ when a sample is sheared to different strains in a given orientation. As the indicator based on each of the six methods are expected to correlate positively with plastic response, i.e., larger indicator means larger $D^2_{min}$, here we define

$$C_{rank} = 2\overline{CDF}(\rho_i) - 1, \qquad [5]$$

where $\rho_i$ is the plastic susceptibility of particle $i$ predicted via one of the six routes, $CDF(\rho_i)$ is the cumulative distribution function value for $\rho_i$, and the bar on top represents averaging $CDF(\rho_i)$ over all particles with top 0.5% $D^2_{min}$. The highest (lowest) value of $C_{rank}$ is around 0.995 (-0.995), which means perfect correlation (anticorrelation), and $C_{rank} = 0$ means no correlation.

As seen in Figs. 3c and d, the $C_{rank}$ achieved with our CNN method is almost equal to its highest possible value over a wide range of strain for both the 2D and 3D glasses. The worst case is $\Omega_a$, for which $C_{rank}$ is always close to zero across the board, for either the 2D or 3D glasses. For the three data-driven methods, their relative predictive power (the magnitude of $C_{rank}$) follows an order similar to that observed in Figs. 3a and b. The CNN prediction stands out to be the best. Note that the predictive power generally decreases with increasing strain for physical indicators. In this regard, our CNN has a major advantage over other methods at larger strains, which is important for monitoring the correlation between structure and mechanical response of amorphous solids. As soft mode analysis on large 3D samples is too computationally expensive, we used instead $Cu_{50}Zr_{50}$ glasses containing 10,000 atoms. Such smaller samples were also analyzed via CNN, to allow head-to-head comparison (included in Fig. 3d). The consistency in $C_{rank}$ obtained with our CNN models on smaller and larger samples (red circles almost overlap with white pentagons in Fig. 3d) indicates the robustness of our CNN models with no sample-size dependence. This is not a big surprise, as the $r_c$ we used is sufficiently large, and so are our training samples.

In addition, we also used a different and extremely stringent criterion, an overlap ratio, $\mathcal{R}$, to gauge the fraction of correctly predicted particles. $\mathcal{R}$ is evaluated by dividing the number of particles belonging simultaneously to two groups that are intersecting (partially overlapping), with the number of particles known to have the top 0.5% $D^2_{min}$. Here the intersection/overlapping particles are those that fall into not only the group known to have the top 0.5% $D^2_{min}$, but at the same time also the group having the top 0.5% plastic susceptibility predicted using that particular method/indicator. In the ideal case, the particles predicted to have the top 0.5% plastic susceptibility are expected to be the same particles known to have the top 0.5% $D^2_{min}$ (i.e., all the particles in that pool). As such, the closer to 1.0 the $\mathcal{R}$ ratio, the higher the predictive capability. The corresponding results are displayed in Supplementary Fig. 15, which shows a similar trend as in Figs. 3c and d. This further confirms the advantage of our CNN method as the most robust of the six, disregarding the metric adopted to characterize the strength of correlation. Under this extremely stringent criterion, the advantage of CNN over GNN is evident on the 2D glasses, see Supplementary Fig. 15a.

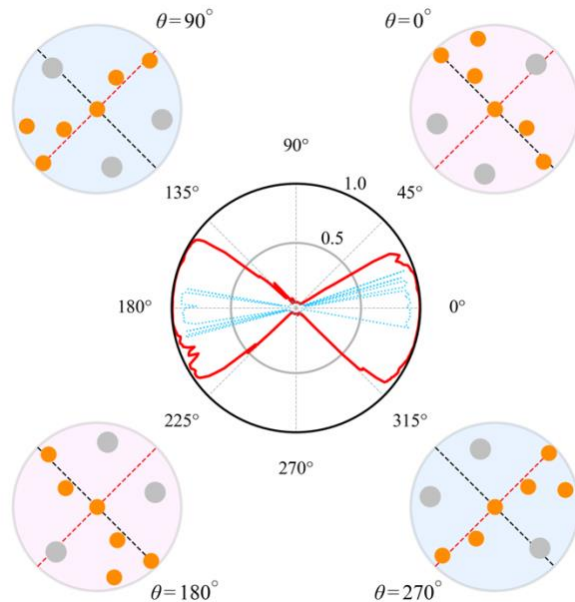

**Fig. 4. The orientation-dependence of CNN-predicted plastic susceptibility for a 2D local configuration.** This susceptibility is characterized here by the magnitude of class probability (red solid curve), varying in the range of 0 to 1 (black circle in the center), depending on the rotation angle ($\theta$) of the local configuration relative to the loading orientation (four scenarios during the counter-clockwise rotation at representative angles are displayed). The red and black dashed lines in each snapshot represent the elongation and contraction directions of the global loading, respectively. The blue dotted line in the polar plot represents $D^2_{min}$ of the local configuration at different $\theta$.

To demonstrate that the orientation dependence has been clearly captured, we show in Fig. 4 the CNN-predicted plastic susceptibility for a 2D local configuration when it is rotated gradually: the rotation-variant SDM turns out to be highly sensitive to minor orientation change. In its initial orientation ($\theta = 0°$), when its loosest and densest packing directions align with the elongation (red dashed line) and contraction (black dashed line) directions of mechanical loading, respectively, CNN predicts the highest plastic susceptibility (class probability very close to 1.0). With the counter-clockwise rotation, these packing directions gradually misalign with the loading ones, and the predicted plastic susceptibility goes down. At $\theta \cong 90°$, the loosest packing of the configuration lines up with the contraction direction whereas the densest packing with the elongation direction, class probability drops to a value very close to zero. When $\theta$ increases to ~180°, the favorable alignment comes back, such that the plastic susceptibility is predicted to be the highest again. Likewise, the lowest reemerges when $\theta = 270°$. This sensitivity to the coupling between the anisotropic local structure and the loading direction demonstrates that, when we feed the images (configurations) into CNN, the algorithm is very good at recognizing this coupling and churning out predictions consistent with the insight in Fig. 2. The blue dotted line in the polar plot of Fig. 4 represents $D^2_{min}$ of the local configuration at different $\theta$ (see *Methods*), which almost matches the CNN prediction. This underscores the benefit and advance enabled by our approach to innovate a new structure representation and combine it with CNN.

## Generalize to different processing history or compositions

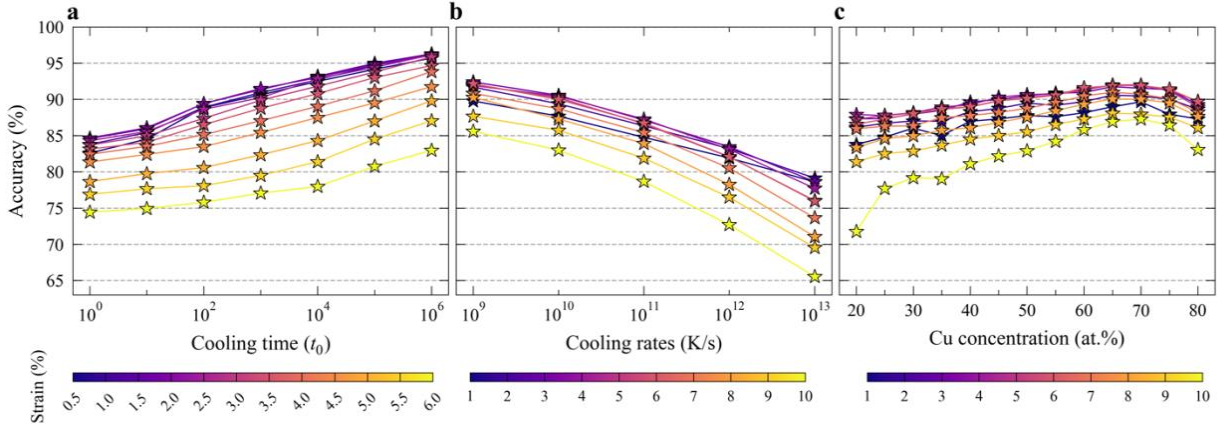

**Fig. 5. The ability of CNN models to generalize**. The CNN models were trained on samples at a single composition with one processing history (the 2D glasses at the composition of $N_L:N_S = (1+\sqrt{5}):4$, quenched within $10^6$ $t_0$. The 3D $Cu_{50}Zr_{50}$ glasses were quenched at $10^{10}$ K/s). The accuracy of the models is shown in (**a**) for 2D glasses quenched with different cooling times, in (**b**) for $Cu_{50}Zr_{50}$ glasses quenched with different cooling rates, and in (**c**) for $Cu_xZr_{100-x}$ glasses of different compositions quenched at a constant effective cooling rate of $10^{10}$ K/s.

Our previous work[30] demonstrated that the SVM models, trained on samples having the same processing history at a single composition, can be generalized to samples with different processing history or at different compositions in the same alloy system. To see if such a generalization is possible with our CNN models, 50 2D samples were quenched over different time periods in the range from $10^0$ to $10^6$ $t_0$. For 3D glasses, we also quenched 10 $Cu_{50}Zr_{50}$ samples with effective cooling rates in the range from $10^9$ to $10^{13}$ K/s, and 10 $Cu_xZr_{100-x}$ glasses with different compositions (x=20, 25, …, 80) at the same effective cooling rate of $10^{10}$ K/s. Each of these 2D (3D) samples was sheared using 4 (24) different loading protocols with AQS method. We then constructed test datasets for each processing history and each composition, by selecting all the atoms with $D^2_{min}$ above a threshold and a same number of atoms randomly from the rest. The accuracy achieved on these datasets, using the CNN model previously trained at a single composition for a particular processing history (see caption and Methods) is presented in Fig. 5a-c. We find a very high accuracy across a wide composition range (Fig. 5c), especially for samples with slower cooling (Figs. 5a and b). For example, for 3D $Cu_{50}Zr_{50}$ quenched at cooling rate of $10^9$ K/s, the highest accuracy achieved is 92.40%. The decrease in accuracy for faster cooling rate was observed before [22, 30, 33, 46], as the critical structural difference becomes increasingly difficult to distinguish[33]. The more relaxed glasses, on the other hand, have a much smaller population of fertile sites that exhibit larger shear susceptibility, making the recognition easier and prediction more robust.

This transferability of the model offers the desirable applicability to glasses across different compositions and/or different processing history, to compare plastic susceptibility. An example is shown in Supplementary Fig. 16**a** (for 2D) and **b** (for 3D). Here the glass structure is generically characterized by the fraction of predicted "$y_i=1$" particles out of all atoms. This fraction, $f_{py=1}$ (the subscript means "predicted (labeled) $y=1$"), is given by the fraction of atoms with class probability > 0.5, and scales almost linearly with the average class probability. These plots reveal a strong correlation between the glass structure and shear modulus ($G$), found to be robust at any strain as long as it is well below that corresponding to global yielding.

A possible reason for the transferability is that our training datasets, although from one composition and the same processing history, are large enough to include almost all of the possible local environments for samples with different processing histories and/or at different compositions within a same alloy systems, as suggested in our previous work[30]. Achieving transferability across different alloys systems while keeping high predictive capability requires more exploration but it is beyond the scope of the current work.

## Conclusion

Our results presented in this paper bring about several advances over previous efforts to identify structural "defects" in amorphous solids, i.e., particles that are most prone to rearrangements, in particular shear transformations. First of all, a qualitative but crucial structural difference, in terms of a more non-uniform and asymmetric packing environment, was discovered for atoms with high plastic susceptibility relative to those with low plastic susceptibility. This difference is found responsible for the anisotropy of local mechanical response in amorphous solids, resolving a puzzle existing for more than 20 years. Our expectation based on this insight, starting from Fig. 2, is reflected in the eventual CNN predictions, as highlighted in Fig. 4. Second, we achieve unprecedented accuracy in predicting orientation-dependent local mechanical response over a wide range of shear strain, by designing a rotation-variant structure representation coupled with state-of-the-art deep learning algorithm, as highlighted in Fig. 3. This unveils the predictive power of the static structure of amorphous solids in forecasting local plastic response. Third, once the optimized CNN model is in hand, all that is needed to predict plastic response is atomic positions, without relying on other knowledge such as interparticle interactions required by previous approaches using parameters based on physical properties. Fourth, we have demonstrated that the CNN models trained on a single glass (with one set of a particular composition and a specific processing history) can be generalized to samples with different processing history or at different compositions in the same alloy system. This is important for probing into the effects of processing procedure or chemical composition on properties, enabling the comparison between different glasses. These four merits open new avenues to the understanding of the structure-property relations in amorphous solids. Finally, we anticipate that our novel structure representation in combination with the powerful CNN method would find use in the studies of other amorphous matter, beyond metallic glasses, as well as in predicting rotation-dependent properties of crystalline materials, via data augmentation[36].

## Methods

The data used for training and validating deep learning models are from molecular dynamics simulations[47], which are implemented in the LAMMPS package[48].

**Preparation of 2D model glasses**. 2D binary glasses are composed of equal-massed ($m$) small ($S$) and large ($L$) particles which interact via a standard 6-12 Lennard-Jones potential. The full details of the potential are presented elsewhere[40]. We chose our composition such that the ratio between the species is $N_L:N_S = (1 + \sqrt{5}):4$ to be consistent with previous studies[14, 22, 32] of this system and each of the sample contains $N$=10,000 particles. All units will be expressed in terms of $m$ as well as $\epsilon$ and $\sigma$, the parameters describing the energy and length scale, respectively, of the interparticle interaction. The characteristic time is $t_0 = \sigma\sqrt{m/\varepsilon}$. The glass transition temperature $T_g$ of this system is known to be 0.325 $\varepsilon/k$[14, 22, 32], where $k$ is the Boltzmann constant. Periodic boundary conditions were imposed on square boxes of linear dimension 98.8045$\sigma$. The density of the system ($N/L^2 \approx 1.02\sigma^{-2}$) was kept constant. 5,000 2D glass samples were each obtained by continuously decreasing the temperature from a liquid state, well-equilibrated at 1.08 $T_g$ and quenched to a low-temperature (0.092 $T_g$) solid state over a period of $10^6$ $t_0$ using a Nose-Hoover thermostat. Then a static relaxation using conjugate gradient algorithm was applied to bring the system to mechanical equilibrium before conducting deformation simulation[22]. Particle positions in this state were used to construct structure representations. Our training, validation and test datasets contained 4,900, 50 and 50 samples, respectively. To demonstrate that the CNN models are valid for the glasses with different processing history, 50 additional samples were quenched from 1.08 $T_g$ to 0.092 $T_g$ over different period in the range from $10^5$ $t_0$ to $10^0$ $t_0$.

**Deforming 2D model glasses.** Each of the 2D glasses was deformed under four loading conditions (positive and negative simple shear, and positive and negative pure shear) with an athermal quasi static (AQS) method[44, 45]. In positive (negative) simple shear, the applied strain increment was $\Delta\gamma_{xy} = \Delta\gamma$ ($\Delta\gamma_{xy} = -\Delta\gamma$), and in positive (negative) pure shear, the applied strain increments are $\Delta\varepsilon_{xx} = -\Delta\varepsilon_{yy} = -\Delta\gamma/2$ ($\Delta\varepsilon_{xx} = -\Delta\varepsilon_{yy} = \Delta\gamma/2$). Here $\Delta\gamma = 10^{-5}$. After each deformation increment, the system was relaxed to its mechanical equilibrium.

**Mechanical response of a local 2D configuration at arbitrary orientation.** To probe the mechanical response of a local 2D configuration at arbitrary orientation, i.e., the blue dotted curve in the Fig. 4, we deleted particles[49] outside the largest circular region from the original square box. Next, we rotated the circular configuration by various angles counter-clockwise, and then conducted simple shear on it with the AQS method. During deformation, the particles in the outer annulus (wall) of width 5 $\sigma$ (two times of potential cutoff) only experience applied affine displacement, i.e., they are not allowed to relax such that the periodic boundary conditions are lost.

**Preparation of 3D model glasses.** The 3D $Cu_{50}Zr_{50}$ metallic glasses were simulated with an optimized embedded atom method (EAM) potential adopted from Ref.[41]. Each of the 1,500 samples contained 32,000 atoms in a cubic box. After adequate equilibration at 2,000 K, the liquid was first quenched to 1,500 K at a cooling rate of $10^{13}$ K/s, followed by $10^{12}$ K/s to 1000 K, then at the desired rate of $10^{10}$ K/s (effective cooling rate, in the text) to 500 K, and finally at $10^{13}$ K/s to 50 K in the NPT ensemble using a Nose-Hoover thermostat with zero external pressure. The periodic boundary condition was applied in all three directions. The samples were then brought to mechanical equilibrium through a static relaxation via a conjugate gradient algorithm. Our training, validation and test datasets for the 3D system contain 1480, 10 and 10 samples, respectively. To demonstrate that the CNN models are valid for glasses with different processing history or at different compositions, 10 $Cu_{50}Zr_{50}$ glasses were also quenched using different effective cooling rates in the range from $10^9$ to $10^{13}$ K/s, and at each different composition ($Cu_xZr_{100-x}$, x=20, 25 ..., 80, in step of 5), 10 MG glasses were quenched with an effective cooling rate of $10^{10}$ K/s.

**Deforming 3D model glasses.** Before extracting atomic coordinates to construct structure representation and conducting deformation simulation in each of the 24 loading orientations listed in Supplementary Table 1, we rotated each configuration such that the shear direction and the normal direction of the shear plane were parallel to X and Y axis, respectively, of the coordinate system. For some loading orientations, appropriate atom replication was needed after rotation, which is illustrated in Supplementary Fig. 17. Each of the 3D glasses was deformed in simple shear along the 24 different loading orientations via an athermal quasi static (AQS) method[44, 45]. After each applied strain increment of $\Delta\gamma_{xy}=10^{-4}$, the system was relaxed to its mechanical equilibrium. We calculated $D^2_{min}$ per definition in Ref.[32], implemented in the OVITO package[50].

**Upper bound of the prediction accuracy.** To assess this ceiling, we make use of the particles from the test dataset, which were already labeled after deformation. Another deformation simulation was then conducted, after shuffling the sequence of all particles, as a new job run on a different computer. The resultant new configuration was used to compute $D^2_{min}$ of all the particles, which were labeled again. The number of particles always having the same label, determined based on both the old and the new $D^2_{min}$, was divided by the total number of particles in the dataset, to give the upper bound of accuracy that can be reached via any method. Note that numerical error, when the atom sequence in the initial configuration file is changed although all other deformation conditions are kept the same, renders deformation under AQS not exactly reproducible. This error also grows with increasing strain.

**Training procedure of CNN models.** For the CNN models, we minimized the binary cross-entropy loss between true labels and predicted labels, and used early stopping and selected the model with the highest accuracy in the validation dataset. The learning rate started from 0.001 and was then divided by $\sqrt{10}$ once the validation accuracy plateaued. We chose RMSprop optimizer and mini-batch size of 512 and 64, for the 2D and 3D glasses, respectively. We augmented the training data by applying randomly one of the 4 (or 8) mirror-symmetries to the images of 2D (or 3D) glasses every time we fed the images into the CNN. The training was implemented in the Tensorflow package[50]. And we conducted distributed deep learning with the Horovod package[52] to accelerate the training for 3D glasses.

**GNN methods**. We followed the procedure in Ref.[31] to construct graph with edge threshold of 2.0 $\sigma$ (and 4.0 Å) for 2D (and 3D) glasses. We used the same GNN architecture as in Ref.[31], except that we activated the output layer with a sigmoid function as we were doing binary classification tasks, and chose $n_{rec}$ = 7 and 6 for 2D and 3D glasses, respectively, according to the optimization results shown in Supplementary Fig. 18. We also tried more neurons in all multilayer perceptors, but this did not further improve the validation accuracy.

For the 2D (3D) samples in both the test and validation datasets, we loaded each sample, and constructed corresponding graph, 4 (24) times, each graph corresponding to one of the 4 (24) loading protocols. And we updated edge vectors of those graphs using relative atomic positions in the new coordinate system, the X and Y axes of which were parallel to the shear direction and the normal of shear plane, respectively, of each specific loading protocol. Atoms were labeled based on the magnitude of $D^2_{min}$. When evaluating the loss function and accuracy for our GNN models, only the atoms in the test and validation datasets were considered. Therefore, both test and validation datasets for the GNN are exact the same as those for our CNN. For the 2D (3D) samples in training datasets, we constructed only one graph for each sample in order to relieve memory burden and save computation time. However, we took into account all loading directions for each of the 4,900 (1,480) training examples by updating at random their target labels with $D^2_{min}$ from one of the 4 (24) loading protocols and by rotating edge vectors correspondingly every time we fed them (both targets and graphs) to the network (note that in our earlier attempts, we tried to directly construct and load multiple graphs for each training sample, but the results obtained were similar). Also, we augmented the training data by applying randomly one of the 4 (or 8) mirror-symmetries to the edge vectors of graphs of 2D (or 3D) glasses every time we fed the graphs to the GNN. When optimizing GNN models and evaluating loss function and accuracy, only the atoms in the training datasets were used to enable balanced datasets and a fair comparison with our CNN.

To minimize the binary cross-entropy loss between true labels and predicted labels, similarly to Ref.[31] we used early stopping and selected the model with the highest accuracy on the validation dataset after running enough epochs. We trained the GNN models with a learning rate of $10^{-4}$, gradient clipping, Adam optimizer and a Tensorflow implementation.

**SVM methods**. Same as in Ref.[28], we used the following radial and angular structural functions to construct input features to SVM models:

$$G_{i,\alpha}(r) = \sum_{j \in \alpha, r_{ij} < r_c} \exp\left(-\frac{(r_{ij}-r)^2}{2\Delta^2}\right),$$

$$\Psi_{i,\alpha,\beta}(\xi,\lambda,\varsigma) = \sum_{j \in \alpha} \sum_{k \in \beta} \exp\left(-\frac{r_{ij}^2 + r_{jk}^2 + r_{ik}^2}{\xi^2}\right)(1 + \lambda \cos \theta_{ijk})^\varsigma,$$

where $r_{ij}$ denotes the distance between atoms $i$ and $j$, and $\alpha$ or $\beta$ denotes the species, $\theta_{ijk}$ is the angle between vector $\boldsymbol{r}_{ij}$ and $\boldsymbol{r}_{ik}$, and $r \in [r_0 + 0.5\Delta, r_c)$ with an increment of $\Delta$. We use $r_0=0.5$ $\sigma$, $r_c=5.0\sigma$ and $\Delta=0.025$ $\sigma$, and $r_0=2.0$ Å, $r_c=8.0$ Å and $\Delta=0.050$ Å for 2D and 3D glasses, respectively, based on the findings in Supplementary Fig. 19. For the parameters in the angular structural functions, similar to Ref.[31], we used $\lambda \in \{-1,1\}, \varsigma \in \{1,2,4,8,16\}$, and 16 values of $\xi$ equally spaced between 0.75 $\sigma$ and 5.0 $\sigma$ (between 3.0 Å and 8.0 Å) for 2D (3D) glasses. We tried to use 4, 8, 16, or 32 values for $\xi$ in the given region but found only a weak dependency. We also tried to use exactly the same angular parameters as in the original work[28], but the results came out slightly worse.

We conducted L2-regularized L2-loss support vector classifications with a linear kernel, which were executed in the LIBLINEAR package[53] by solving the primal optimization problem. A regularization parameter of $C = 100$ was used, as we tried many different values over the range from $10^{-7}$ to $10^5$ and found only a weak dependency once $C$ was above 10. We also tuned the termination criterion, which did not further improve the accuracy on validation datasets. Note that 500 (100) samples were sufficient to construct training datasets for 2D (3D) glasses, because training the SVM models does not require large datasets, as demonstrated in Supplementary Fig. 20.

Supplementary Table 2 and Supplementary Table 3, for 2D and 3D glasses, respectively, present a summary of the input features and trainable parameters of the three machine learning models.

**Calculating the flexibility volume**. The flexibility volume is defined as the product of vibrational mean squared displacement (MSD) and atom spacing[19]. We calculated MSD at 0.092 $T_g$ for 2D glasses and at 50 K for 3D glasses, using the procedure in Ref.[19]. The MSD was averaged over 100 independent runs, all starting from the same configuration but different initial velocities. And the atomic volume is calculated based on Voronoi analysis.

**Soft mode analysis**. The normal mode analysis of the glasses was carried out by diagonalizing the dynamical matrix of the inherent structure of glasses. The participation fraction of atom $i$ in eigenmode $e_\omega$ is $p_i = \left|\vec{e}_\omega^i\right|^2$, where $\vec{e}_\omega^i$ is the corresponding polarization vector of atom $i$[54]. Here, $p_i$ was summed over a small fraction of 1% (same as that in Ref.[18]) of the lowest-frequency normal modes. $p_i$ measures the involvement of atom $i$ in soft modes.

**Data availability**

The data that support the findings of this study are available from the corresponding author on request.

**Acknowledgements**

The authors gratefully acknowledge Prof. M. L. Falk for discussions and his insightful comments on the manuscript, and V. Bapst for the useful discussions on training the GNN model. The work is supported at JHU by U. S. Department of Energy (DOE) DOE-BES-DMSE under grant DE-FG02-19ER46056. This research used the CPU resources of the National Energy Research Scientific Computing Center (NERSC), a DOE Office of Science User Facility supported by the Office of Science of the U.S. Department of Energy under Contract No. DE-AC02-05CH11231, and the Maryland Advanced Research Computing Center (MARCC). The authors also acknowledge the Texas Advanced Computing Center (TACC) at The University of Texas at Austin for providing GPU resources that have contributed to the research results reported within this paper. The early attempts of deep learning used the GPU resource from MARCC.

**Author contributions**

Z.F. and E.M. conceived the research project. Z.F. designed the structure representation, conducted the MD simulations and carried out the machine/deep learning analysis. Both authors contributed to the discussions and the writing of the paper. E.M. left the project in July 2020.

**Competing interests**

The authors declare no competing interests.

**Additional information**

**Supplementary information** is available for this paper at xxx.